\documentclass{iopart}
\usepackage[dvips]{graphicx}
\usepackage{amssymb}
\usepackage{hyperref}
\newcommand{\abs}[1]{\left|#1\right|}
\newcommand{\text}[1]{\hbox{\scriptsize\rm #1}}
\newcommand{\un}[1]{\mathrm{\,#1}}

\newcommand{\PRpt}{{\it Phys. Rep.}}
\newcommand{\AandA}{{\it Astron. Astrophys.}}
\begin{document}
\title[First LIGO Science Results: Stochastic Background Upper Limits]
{First upper limit analysis and results from LIGO science data:
  stochastic background\footnote{
    Published as \CQG\ \textbf{21} (2004) S685-S690;\\ available online at
    \href{http://stacks.iop.org/CQG/21/S685}{stacks.iop.org/CQG/21/S685};
    copyright (C) 2004 IOP Publishing Ltd.
  }}
\author{John T Whelan (for the LIGO Scientific
  Collaboration\footnote{The members of the LIGO Scientific
    Collaboration are listed in the paper by Allen and Woan for the
    LIGO Scientific Collaboration, in these proceedings:
    \href{http://stacks.iop.org/CQG/21/S671}{stacks.iop.org/CQG/21/S671}})}
\address{Department of Physics, Loyola University, New Orleans,
  Louisiana 70118, USA}
\begin{abstract}
  I describe analysis of correlations in the outputs of the three
  LIGO interferometers from LIGO's first science run, held over 17
  days in August and September of 2002, and the resulting upper limit
  set on a stochastic background of gravitational waves.  By searching
  for cross-correlations between the LIGO detectors in Livingston, LA
  and Hanford, WA, we are able to set a 90\% confidence level upper
  limit of $h_{100}^2\Omega_0<23\pm4.6$.
\end{abstract}
\ead{jtwhelan@loyno.edu}
\maketitle

\section{Introduction}
\label{s:intro}

The LIGO interferometric gravitational wave (GW) detector held its
first science run (S1) in 2002, from 23 August to 9 September
\cite{LIGOS1}.  LIGO consists of an interferometer (IFO) with 4~km
arms in Livingston, LA, USA (the LIGO Livingston Observatory, or LLO),
called L1 for short, and two IFOs, with arms of 4~km and
2~km, in Hanford, WA, USA (the LIGO Hanford Observatory, or LHO),
called H1 and H2, respectively.  The data were
analysed to search for GW bursts \cite{S1burst}, signals from
inspiralling neutron star binaries \cite{S1inspiral}, periodic signals
from a rotating neutron star \cite{S1pulsar}, and stochastic
backgrounds \cite{S1stoch}.  This paper summarizes the analysis method
and results of the search for a stochastic background of gravitational
waves (SBGW), which are explained in more detail in \cite{S1stoch}.

\section{Fundamentals of Analysis Method}
\label{s:method}

A SBGW is assumed for simplicity to be  isotropic, unpolarized, Gaussian,
and stationary.  Subject to these assumptions, the SBGW
is completely described by its power
spectrum.  It is conventional to express this spectrum in terms of the
GW contribution to the cosmological parameter
$\Omega=\rho/\rho_{\text{crit}}$:
\begin{equation}
  \label{eq:omegagw}
  \Omega_{\text{GW}}(f)=\frac{1}{\rho_{\text{crit}}}
  \frac{d\rho_{\text{GW}}}{d\ln f}=\frac{f}{\rho_{\text{crit}}}
  \frac{d\rho_{\text{GW}}}{df}
  \ .
\end{equation}
Note that $\Omega_{\text{GW}}(f)$ has been constructed to be
dimensionless, and represents the contribution to the overall
$\Omega_{\text{GW}}$ per \emph{logarithmic} frequency interval.  In
particular, it is \emph{not} equivalent to $d\Omega_{\text{GW}}/df$.
Note also that since the critical density $\rho_{\text{crit}}$, which
is used in the normalization of $\Omega_{\text{GW}}(f)$, is
proportional to the square of the Hubble constant $H_0$ \cite{Kolb:1990},
it is convenient to work with $h_{100}^2\Omega_{\text{GW}}(f)$, which
is independent of the observationally determined value of
$h_{100}=\frac{H_0}{100 \un{km}/\un{s}/\un{Mpc}}$.\footnote{Although
  $h_{100}$ is now much more accurately known than it once was, we
  still work in terms of $h_{100}^2\Omega_{\text{GW}}(f)$ to
  facilitate comparisons with prior results.}

The standard method to search for such a background is to
cross-correlate the outputs of two GW detectors
\cite{Christensen:1992}.  If the noise in the two detectors is
uncorrelated, the only non-zero contribution to the average
cross-correlation (CC) will come from the SBGW.  In the
optimally-filtered CC method (described in more detail
in \cite{Allen:1997,Allen:1999,Whelan:2001}), one calculates a
CC statistic
\begin{equation}
  Y
  =
  \int dt_1\, dt_2\, {h_1(t_1)}\, {Q(t_1-t_2)}
  {h_2(t_2)}
  \label{eq:CCstat}
  =
  \int df\,{\widetilde{h}_1^*(f)}\, {\widetilde{Q}(f)}\,
  {\widetilde{h}_2(f)}
\end{equation}                                
where $h_{1,2}(t)$ are the data streams from the two detectors,
$\widetilde{h}_{1,2}(f)$ are their Fourier transforms, and $Q(t_1-t_2)$ (with
Fourier transform $\widetilde{Q}(f)$) is a suitably-chosen optimal
filter.  The choice which optimizes signal-to-noise ratio for a
constant-$\Omega_{\text{GW}}(f)$ background is \cite{Allen:1999}
\begin{equation}
  \label{eq:Qdef}
  \widetilde{Q}(f)\propto\frac{\gamma(f)}{f^3P_1(f)P_2(f)}
\end{equation}
The normalization of the optimal filter is conventionally chosen so
that in the presence of a SBGW of strength
$\Omega_{\text{GW}}(f)=\Omega_0$, the expected mean value of the
CC statistic is
\begin{equation}
  \label{eq:CCmean}
  \langle Y \rangle = h_{100}^2\Omega_0 T
\end{equation}
where $T$ is the duration of the analysed data sets.  The expected
variance of the CC statistic is
\begin{equation}
  \label{eq:CCvar}
  \sigma_{\text{theor}}^2
  = \frac{T}{4} \int df\,P_1(f)\,\abs{\widetilde{Q}(f)}^2\,P_2(f)
  \propto 
  \left(
    \int \frac{df}{f^6} \frac{[\gamma(f)]^2}{P_1(f)P_2(f)}
  \right)^{-1}
  \ .
\end{equation}
The method is sensitive to backgrounds on the
order of
\begin{equation}
\label{eq:CCsens}
\Omega^{\text{UL}}
\sim
\left(
  {T}
  \int df\frac{[\gamma(f)]^2}{f^6 {P_1(f) P_2(f)}}
  \right)^{-1/2}
  \ .
\end{equation}
The sensitivity of this method improves with time and is limited by the power
spectral densities $P_{1,2}(f)$ of the noise in the two detectors.
The factor
\begin{equation}
  \gamma(f)={d_{1ab}}\ {d_2^{cd}}\
  {{5\over 4\pi}\int_{S^2}}{d^2\Omega}\
  {e^{i2\pi f{\mathbf{n}}{\cdot}
      {\mathbf{\Delta x}}{/c}}}\
  {P^{ab}_{cd}(\mathbf{n})}
\end{equation}
in the numerator of the integral is the \emph{overlap reduction
  function} \cite{Flanagan:1993}, which describes the observing
geometry.  Here $P^{ab}_{cd}(\mathbf{n})$ is a projector onto
symmetric traceless tensors transverse to a direction $\mathbf{n}$
and $d_{1,2}^{ab}$ are the \textit{detector response tensors} for the
two detectors.  These are the tensors with which the metric
perturbation $h_{ab}$ at the detector should be contracted to obtain
the \textit{GW strain} $h=d^{ab}h_{ab}$.  If $u_a$ and
$v_a$ are unit vectors pointing in the directions of an
IFO's two arms, its response tensor is
\begin{equation}
  d_{ab} = \frac{1}{2}(u_a u_b - v_a v_b)
  \ .
\end{equation}

\begin{figure}[htbp]
  \vspace{5pt}
  \begin{center}
    \includegraphics[height=3.5in]{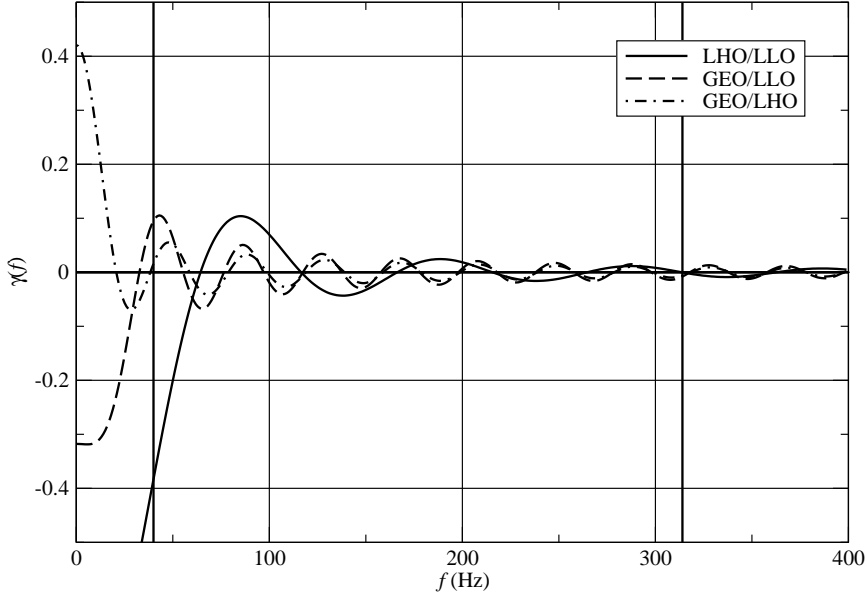}
  \end{center}
  \caption{
    The overlap reduction function $\gamma(f)$ for combinations of the
    LIGO Livingston Observatory (LLO) and LIGO Hanford Observatory
    (LHO) with each other and with the GEO600 site.  (GEO600 was
    also operational during S1, but has not been included in this
    analysis because it was considerably less sensitive than the LIGO
    detectors.)  Note that the overlap reduction function for
    correlations between the two detectors at LHO (H1 and H2) is
    identically equal to unity.  The solid lines (at $f=40\un{Hz}$ and
    $314\un{Hz}$) show the range of frequencies used in our analysis.
  }
  \label{fig:alloverlap}
\end{figure}

The overlap reduction function is equal to unity for the case of a
pair of IFOs at the same location with their
arms aligned, and is suppressed as the detectors are rotated out of
alignment or separated from one another.  It also oscillates with
frequency as correlations are suppressed for detectors whose
separation is comparable to or greater than the corresponding GW
wavelength.  Figure~\ref{fig:alloverlap}
shows the overlap reduction functions for combinations of detectors
which were operational during S1.

\section{Prior Results}
\label{s:prior}

The previous best upper limit on a SBGW from direct observation with GW
detectors was $h_{100}^2\Omega_{\text{GW}}(900\un{Hz})\le 60$
\cite{Astone:1999}, set by correlating the resonant bar detectors
Explorer (in Geneva, Switzerland) and
Nautilus (near Rome, Italy). 
A broad-band limit of $h_{100}^2\Omega_{\text{GW}}(f)\le
3\times 10^5$ was set using a pair of `prototype' IFOs
\cite{Compton:1994}.

More stringent upper limits can be set on astrophysical grounds.  They
are detailed elsewhere \cite{S1stoch,Allen:1997,Maggiore:2000}, but we
mention the
bound from big-bang nu\-cle\-o\-syn\-the\-sis \cite{Kolb:1990,Maggiore:2000},
which states that a cosmological SBGW is limited by
\begin{equation}
  \int_{10^{-8}\un{Hz}}^\infty \frac{df}{f}h_{100}^2\Omega_{\text{GW}}(f) 
  \le 10^{-5}
  \ .
\end{equation}
This broad-band limit implies that any cosmologically interesting SBGW
very likely lies several orders of magnitude below the existing
limits.

\section{Details of Analysis Method}
\label{s:details}

Each of the three combinations of detectors (H1-L1, H2-L1, and H1-H2)
was analysed separately for CCs.  Since the power spectra
$P_{1,2}(f)$ varied over the course of the S1 run, the co\"{\i}ncident
data for each pair of IFOs were divided into 15~min
blocks, and an optimal filter constructed for each such
block.\footnote{The power spectra were constructed using Welch's
  method, with 449 overlapping Hann-windowed periodograms averaged to
  produce a power spectrum estimate with a resolution of
  $0.25\un{Hz}$.}  To maximize overall signal-to-noise
ratio \cite{Allen:1999}, we combined the CC statistics from the
different blocks using a weighting factor of
$\sigma_{\text{theor}}^{-2}$, where $\sigma_{\text{theor}}$ is the
theoretical standard deviation defined in (\ref{eq:CCvar}).  Note that
this can be calculated from the individual power spectra, without
cross-correlating the data.  To avoid problems from noisy and
presumably less Gaussian data, we discarded the 15~min blocks
with the highest $\sigma_{\text{theor}}$ values, corresponding to a
5\% total contribution to $\sum\sigma_{\text{theor}}^{-2}$.

For each block, an optimal filter was constructed with a frequency
resolution of $0.25\un{Hz}$ according to the discrete frequency-domain
analogue of (\ref{eq:Qdef}).  The range of frequencies included in the
calculation of the CC statistic was chosen to be 40--314\,Hz for H1-L1
and H2-L1, and 40--300\,Hz for H1-H2.  Given the power spectra
of the instruments and the expected spectrum of correlations
associated with a constant $\Omega_{\text{GW}}(f)$,
frequencies outside that range were not expected to improve the
sensitivity appreciably.  Additionally, individual frequency bins
associated with cross-correlated instrumental noise were omitted from
the sum over frequencies (which means the optimal filter was
effectively set to zero there).  These were integer multiples of
16\,Hz and 60\,Hz, as well as a few frequencies which had a coherence
over the entire run above 0.2, namely 250\,Hz for L1-H2, and
168.25\,Hz and 168.5\,Hz for H1-H2.

Each 15~min block was divided into ten 90~s segments using a
Tukey window which consisted of half~s Hann transitions on either
side of an 89~s flat top.  The CC statistic was calculated for
each segment, using the discrete analogue of the
frequency-domain form of (\ref{eq:CCstat}), and these were
combined to give a CC statistic for the entire
block.  In this way, we were able to obtain, through the measured
standard deviation of the ten CC statistics within a block, a
statistical measure of the error associated with the CC statistic for
the block.  We also estimated the systematic error associated with the
change in sensitivity and calibration over the course of each block.\footnote
{
  The calibration of the LIGO detectors was monitored by tracking the
  amplitude of a sinusoidal `calibration line' signal injected
  into the IFO arm length.  The output amplitude, recorded
  once per minute, allows construction of a frequency-dependent
  response function which accounts for IFO alignment
  drifts, as detailed in \cite{LIGOS1,Adhikari:2003}.
}
We then combined all three, appropriately weighted over the whole run,
to obtain a total estimated error $\widehat{\sigma}_{\text{tot}}$
associated with the point estimate $h_{100}^2\widehat{\Omega}_0$
calculated from the weighted average of all the CC statistics using
(\ref{eq:CCmean}).  In the absence of cross-correlated noise, the 90\%
confidence level upper limit on $\Omega_0$, the constant value of
$\Omega_{\text{GW}}(f)$, is
\begin{equation}
  h_{100}^2\Omega_0\le h_{100}^2\widehat{\Omega}_0
  +1.28\widehat{\sigma}_{\text{tot}}
  \ .
\end{equation}

\section{Results}
\label{s:results}

The results for the three IFO pairs are summarized in
Table~\ref{tab:results}.
\begin{table}[htbp]
  \begin{center}
    \begin{tabular}{lcccc}
      \hline
      IFO Pair & {obs time (h:min)} & {$h_{100}^2\widehat{\Omega}_0$} 
      & {$\widehat\sigma_{\rm tot}$} & {90\% CL UL} \\
      \hline
      {H2}-{L1} & {51:15} & {0.2}
      & {18} & {23} \\
      {H1}-{L1} & {64:00} & {32}
      & {18} & {55} \\
      {H1}-{H2} & {100:15} & {-8.3}
      & {0.9} & {N/A} \\
      \hline
    \end{tabular}
  \end{center}
  \caption{Summary of the point estimate $h_{100}^2\widehat{\Omega}_0$
    and total estimated error $\widehat\sigma_{\rm tot}$ for the three
    IFO pairs considered.  Note that no upper limit is set from H1-H2,
    the two IFOs at the Hanford site, since there was evidence of
    cross-correlated noise.  Associated with each of the values quoted
    is an additional 20\% uncertainty arising from the calibration of
    the instruments.}
  \label{tab:results}
\end{table}
There is a statistically significant anti-correlation observed between
H1 and H2, two IFOs which share the same vacuum envelope at the LHO
site in Hanford, WA.  Time-shift and $\chi^2$ analyses show that this
anti-correlation is inconsistent with constant-$\Omega_{GW}(f)$ SBGW,
so we conclude it is due to instrumental cross-correlations between
the two colocated detectors.\footnote{Of course, the fact that it's
  an anti-correlation rather than a correlation is another reason it
  can't be due to a SBGW.}  For the inter-site measurements (H1-L1 and
H2-L1), the lack of statistically significant cross-correlations makes
these checks trivial, and we proceed to setting an upper limit from
each pair.\footnote{One might try to combine the different measurements
  into a single limit, but we choose to consider them individually,
  especially given the observed correlations
  between the two Hanford IFOs, which complicate the issue of
  combining the H1-L1 and H2-L1 as supposedly independent
  measurements.}  The stronger upper limit is set by H2 and L1, and it
is
\begin{equation}
  h_{100}^2\Omega_0\le 23\pm 4.6
  \ .
\end{equation}
This represents a factor of 2--3 improvement over the previous direct
upper limits described in Sec.~\ref{s:prior}, and an improvement by a
factor of over 1000 over the previous measurements with
interferometric detectors.

For more details on the analysis, the reader is directed to
\cite{S1stoch}.

\ack

The authors gratefully acknowledge the support of the United States
National Science Foundation for the construction and operation of the
LIGO Laboratory and the Particle Physics and Astronomy Research
Council of the United Kingdom, the Max-Planck-Society and the State of
Niedersachsen/Germany for support of the construction and operation of
the GEO600 detector. The authors also gratefully acknowledge the
support of the research by these agencies and by the Australian
Research Council, the Natural Sciences and Engineering Research
Council of Canada, the Council of Scientific and Industrial Research
of India, the Department of Science and Technology of India, the
Spanish Ministerio de Ciencia y Tecnologia, the John Simon Guggenheim
Foundation, the David and Lucile Packard Foundation, the Research
Corporation, and the Alfred P. Sloan Foundation.

JTW also wishes to thank his colleagues in the LIGO Scientific
Collaboration, especially the Stochastic Sources Upper Limits Group,
chaired by J~Romano and P~Fritschel, as well as the Albert Einstein
Institute in Golm, Germany.  JTW was supported by the National
Science Foundation under grant PHY-0300609 and by the
Max-Planck-Society.


\section*{References}
\begin{thebibliography}{19}
\bibitem{LIGOS1} Abbot~B et al (LIGO Scientific Collaboration) 2003
  Detector description and performance for the first coincidence
  observations between LIGO and GEO {\it Preprint} gr-qc/0308043 ({\it
    Nucl.\ Instrum.\ Methods} at press)
\bibitem{S1burst} Abbot~B et al (LIGO Scientific Collaboration) 2003
  First upper limits on gravitational wave bursts from LIGO
  {\it Preprint} gr-qc/0312056
\bibitem{S1inspiral} Abbot~B et al (LIGO Scientific Collaboration) 2003
  Analysis of LIGO Data for Gravitational Waves from Binary Neutron
  Stars {\it Preprint} gr-qc/0308069 (\PR D submitted)
\bibitem{S1pulsar} Abbot~B et al (LIGO Scientific Collaboration) 2004
  \PR D {\bf 64} at press ({\it Preprint} gr-qc/0308050)
\bibitem{S1stoch} Abbot~B et al (LIGO Scientific Collaboration) 2003
  Analysis of First LIGO Science Data for Stochastic Gravitational
  Waves {\it Preprint} gr-qc/0312088 (\PR D to be submitted)
\bibitem{Kolb:1990} Kolb E W and Turner M S 1990 {\it The Early Universe}
  (Addison-Wesley, 1990)
\bibitem{Christensen:1992}
  Christensen N 1992 \PR D {\bf 46} 5250
\bibitem{Allen:1997}
  Allen B 1997
  in \textit{Proceedings of the Les Houches School on Astrophysical Sources of
  Gravitational Waves},
  eds Marck J A and Lasota J P, Cambridge, 373;
  \nonum Allen B 1996 {\it Preprint} gr-qc/9604033
\bibitem{Allen:1999}
  Allen B and Romano J D 1999 \PR D {\bf 59} 102001;
  \nonum Allen B and Romano J D 1997 {\it Preprint} gr-qc/9710117
\bibitem{Whelan:2001}
  Whelan J T et al 2001 \CQG  {\bf 19} 1521;
  \nonum Whelan J T et al 2001 {\it Preprint} gr-qc/0110019
\bibitem{Flanagan:1993}
  Flanagan \'{E} \'{E} 1993 \PR D {\bf 48} 2389;
  \nonum Flanagan \'{E} \'{E} 1993 {\it Preprint} astro-ph/9305029
\bibitem{Astone:1999} Astone P et al 1999 \AandA {\bf 351} 811;
  \nonum Astone P et al 1998 in \textit{Proceedings of the
    Second Edoardo Amaldi  Conference on Gravitational Waves},
  eds Coccia E, Pizzella G and Veneziano G, Singapore, 192
\bibitem{Compton:1994}
  Compton C, Nicholson D and Schutz B F 1994
  in \textit{Proceedings of the Seventh Marcel Grossman Meeting on
    General Relativity},
  eds Jantzen R T and Keiser G M, Singapore, 1078
\bibitem{Maggiore:2000} Maggiore M 2000 \PRpt {\bf 331} 28;
  \nonum Maggiore M 1999 {\it Preprint} gr-qc/9909001
\bibitem{Adhikari:2003} Adhikari R, Gonz\'{a}lez G, Landry M and
  O'Reilly B (for the LIGO Scientific Collaboration) 2003
  \CQG {\bf 20} S903
\end{thebibliography}
\end{document}